\documentclass{aa}
\usepackage{graphics}
\usepackage{psfig}

\begin{document}

\title{Kinematics and binaries in young stellar aggregates. \\II. NGC 6913$\equiv$M29}

\author{C. Boeche\inst{1}
\and   U. Munari\inst{1}
\and   L. Tomasella\inst{1}
\and   R. Barbon\inst{2}}

\offprints{U. Munari (munari@pd.astro.it)}

\institute{
Osservatorio Astronomico di Padova, Sede di Asiago,
I-36012 Asiago (VI), Italy
\and
Osservatorio Astrofisico del Dipartimento di Astronomia,
Universit\'a di Padova, I-36012 Asiago (VI), Italy
          }

   \date{Received date..............; accepted date................}

\abstract{
Between 1996 and 2003 we have obtained 226 high resolution spectra of 16
stars in the field of the young open cluster NGC~6913, to the aim of
constraining its main properties and study its internal kinematics. Twelve
of the program stars turned out to be members, one of them probably unbound.
Nine are binaries (one eclipsing and another double lined) and for seven of
them the observations allowed to derive the orbital elements. All but two of
the nine discovered binaries are cluster members. In spite of the young
age (a few Myr), the cluster already shows signs that could be interpreted
as evidence of dynamical relaxation and mass segregation. However, they may
be also the result of an unconventional formation scenario. The dynamical
(virial) mass as estimated from the radial velocity dispersion is larger
than the cluster luminous mass, which may be explained by a combination of
the optically thick interstellar cloud that occults part of the cluster, the
unbound state or undetected very wide binary orbit of some of the
members that inflate the velocity dispersion and a high inclination for the
axis of a possible cluster angular momentum. All discovered binaries are
hard enough to survive average close encounters within the cluster and do
not yet show sign of relaxation of the orbital elements to values typical of
field binaries.
}

\maketitle

	\section{Introduction}

This is the second paper of a series devoted to the results of a long term,
high resolution spectroscopic study of early type members of young open
clusters, trapezium systems and OB associations. The aims of this series are
discussed in Paper~I (Munari and Tomasella 1999).

NGC~6913, the topic of this paper, is a young open cluster harboring O-type
members and lying close to the plane of the Galaxy
($\alpha=20^{h}23^{m}_{\cdot}9$, $\delta=+38^{\circ}32^{\prime}$ (J2000);
$l=76^{\circ}_{\cdot}92$, $b=+0^{\circ}_{\cdot}61$). Despite appearing in
the Messier catalog as M29, few papers in literature deal with it,
furthermore showing some disagreement in the results. Cluster distance is
reported to be 2.2 kpc by Morgan and Harris (\cite{morgan_harris}) and
Massey et al. (\cite{massey}), 1.5 kpc by Joshi et al. (\cite{joshi}), and
1.1 kpc by Hoag et al. (\cite{hoag}), while Tifft (\cite{tifft}) suggested
that NGC~6913 is indeed the results of two separate groups of stars, one at
1.6 kpc and the other somewhere between 1.9 and 2.4 kpc. The mean and
differential reddening span a range of values too: $<E_{B-V}>=$0.78, $\Delta
E_{B-V}$=0.64 according to Joshi et al. (\cite{joshi}), $<E_{B-V}>$=0.71 and
$\Delta E_{B-V}$=1.82 for Wang and Hu (\cite{wang}), and $<E_{B-V}>$=1.03
following Massey et al. (\cite{massey}). Similarly, estimated ages span from
0.3$-$1.75 Myr of Joshi et al. (\cite{joshi}) to 10 Myr of Lyng\aa\
(\cite{linga}).

The internal and galactic kinematics of NGC~6913 has not been so far
investigated in literature. The cluster radial velocity used by Hron (1987)
in modeling the rotation curve of the Galaxy, $-$25~km~sec$^{-1}$, was
assembled by scanty literature data that apparently missed all brightest
cluster members, and is largely off our much more accurate and
representative $-$16.9($\pm$0.6)~km~sec$^{-1}$ value (see sect.~3.2).
Internal kinematics and binary content of NGC~6913 are unknown because no
detailed radial velocity study of its members has been ever pursued, and
proper motions investigations (Sanders \cite{sanders}, Dias et al.
\cite{dias}) are not deep and accurate enough for a firm membership
segregation over a wide range of magnitudes, do not cover all candidate
members and do not allow resolution of the internal kinematics.

In this paper we aim to look in more details to NGC~6913 general properties
(like astrometric membership, photometry, reddening, distance, mass and age)
and to present and discuss the results of our extensive spectroscopic study
of NGC~6913 based on 226 high resolution spectra monitoring of 16 stars in
the field of the cluster over the time span 1996-2003. These observations
are used to constrain the internal velocity dispersion, the cluster
galactic motion, the individual rotational velocities, and the internal
kinematical and evolutionary status of the cluster. Spectroscopic
orbits are calculated for the discovered binary stars.

\begin{table*}[!Ht]
\begin{center}
\caption[]{Program stars. The first four columns give our identification
number (cf. finding chart in Figure~1), and that assigned by Hoag et al.
(1961), Sanders (1973) and Kazlauskas and Jasevicius (1986). $V$ and $B-V$
are Tycho-2 $V_T$ and $(B-V)_T$ transformed into Johnson system following
Bessell (2000) prescriptions. $U-B$ is the median of the measurements by
Massey et al. (\cite{massey}), Joshi et al. (1983) and Hoag et al. (1961).
Star \#10 is reported as a short period variable by Pe\~{n}a et al. 2001.}
\begin{tabular}{rrrrccrcrlcc}
\hline
&&&&&&&\\[-5pt]
\#  & H61 & S73 & KJ86 & HD     & HIP/TYC                   & \multicolumn{1}{c}{$V$}  & $B-V$& $U-B$ & notes\\
&&&&&&&\\[-5pt]
\hline
&&&&&&&\\[-5pt]
      1   & 1 & 135 & 125 & 194378 & HIP 100586                & 8.603  & 0.431    & +0.07 & V2031 Cyg   \\ 
      2   & 2 & 159 & 145 & 229239 & HIP 100612                & 9.035  & 0.730    &--0.14 &    \\
      3   & 3 & 157 & 144 & 229238 & TYC 3152 1325 1           & 8.935  & 0.801    &--0.07 &    \\
      4   & 4 & 149 & 138 & 229234 & TYC 3152 1369 1           & 8.979  & 0.638    &--0.20 &    \\
      5   & 5 & 125 & 118 & 229221 & TYC 3152 1451 1           & 9.260  & 0.767    &--0.25 & V1322 Cyg   \\
      6   & 6 & 139 & 127 & 229227 & HIP 100600                & 9.419  & 0.632    &--0.18 &    \\
      7   & 7 & 174 & 156 & 229253 & TYC 3152 \phantom{1}236 1 & 10.171 & 0.099    &--0.31 &    \\
      8   & 8 & 147 & 136 &        & TYC 3152 1309 1           & 10.388 & 0.733    &--0.17 &    \\
      9   & 9 & 146 & 134 & 229233 & TYC 3152 1137 1           & 10.494 & 0.346    & +0.02 &    \\
     10  &10  & 182 & 162 & 229261 & TYC 3152 1415 1           & 10.510 & 0.252    &--0.31 &  var  \\
     11  &11  & 178 & 158 &        & TYC 3152 1019 1           & 11.307 & 0.518    & +0.26 &    \\
     12  &12  & 122 & 115 &        & TYC 3152 \phantom{1}676 1 & 12.091 & 0.025    &--0.06 &    \\
     13  &13  & 167 & 150 &        & TYC 3152 1467 1           & 11.692 & 0.862    & +0.83 &    \\
     14  &14  & 148 & 137 &        & TYC 3152 1423 1           & 11.552 & 0.426    & +0.04 &    \\
     15  &    & 143 & 132 &        & TYC 3152 \phantom{10}54 1 & 11.534 & 1.120    & +0.28 &    \\
     16  &    &     & 103 &        & TYC 3152 1453 1           & 10.983 & 0.528    &--0.30 &    \\
&&&&&&&\\[-5pt]
\hline
\end{tabular}
\end{center}
\label{prog_stars}
\end{table*}

\begin{figure}[!Ht]
\centerline{\psfig{file=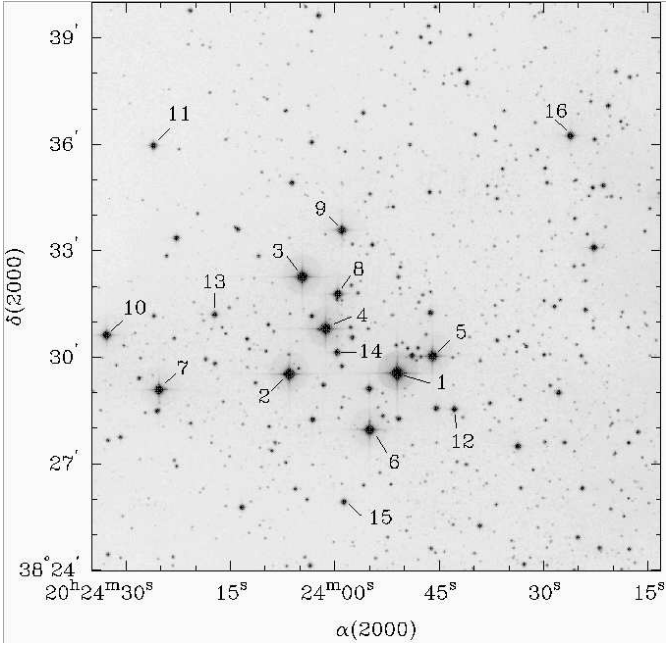,width=8.8cm}}
\caption[]{Finding chart for NGC~6913 program stars.}
\end{figure}

	\section{Spectroscopic observations}

Table~1 summarizes the main properties of the 16 selected program stars, and
Figure~1 provides a finding chart for them. The program stars have been
spectroscopically observed over the period 1996-2003 with the 1.82 cm
telescope and Echelle+CCD spectrograph of the Astronomical Observatory of
Padova at Asiago (Italy). Table~2 provides the journal of observations. The
instrumental set-up, spectra extraction and calibration, accuracies, etc.
are identical to Paper~I and the reader is referred to it for details.

        \subsection{Spectral classification and radial velocities}

Scanty information exists on the spectral classification of the program
stars. Wang and Hu (\cite{wang}) derived spectral types from low resolution
spectra (5.3~\AA/pix) covering the range 4200-6900~\AA. Kazlauskas and
Jasevicius (1986) obtained photoelectric photometry in the Vilnius system,
that we have converted into spectral types using the reddening free color
parameters $Q_{\rm }$ defined by Stray\v{zis} (1977) appropriate for the
$R_{\rm V}$ = $A_{\rm V} / E_{B-V}$=3.6 reddening law that applies to
NGC~6913 according to Johnson~(1962).  We have also derived spectral
classification of the program stars using our Echelle spectra, classified
against the Yamashita et al. (1977) spectral atlas. Even if spectral
classification of Echelle spectra has to be carried out with care (lines to
be compared normally fall on different Echelle orders), nevertheless the
resulting spectral types look quite reasonable, and, given the far superior
spectral resolution and high S/N, also possibly more accurate than those of
Wang and Hu (\cite{wang}). The three estimates of the spectral type are
compared in Table~3. The last two columns of the table give the reddening
and distance when Fitzgerald (1970) intrinsic colors and our spectral
classification are compared to $V$, $B-V$ photometry in Table~1. The
positions of the program stars on the reddening corrected HR diagram are
shown in Figure~2.

\begin{table}[!Ht]
\caption[]{Journal of observations. $D$ is the dispersions (\AA/pix) at
H$\alpha$ (0.19 corresponding to unbinned spectra, 0.38 to 2$\times$ binned
spectra), and $\Delta \lambda$ is the wavelength coverage.
The last column gives the program stars observed in each given run.}
\begin{tabular}{cccl}
\hline
&&&\\[-4pt]
date& D     & $\Delta \lambda$ (\AA)& star \#\\
&&&\\[-4pt]
\hline
&&&\\ [-4pt]
1996.06.28 & 0.38  & 4100-6700 & 1,2,4,5,6,7,10,12,14\\
1996.06.29 & 0.38  & 4100-6700 & 3,8,9,11,13\\
1996.08.01 & 0.38  & 4100-6700 & 1,2,3,4,8,9\\
1996.08.02 & 0.38  & 4100-6700 & 5,6,10,11,12\\
1996.08.03 & 0.38  & 4100-6700 & 1,2,3,4,5,8,9\\
1996.08.04 & 0.38  & 4100-6700 & 6,10,11,12\\
1996.08.20 & 0.38  & 4100-6700 & 3,6,7,10,11\\
1996.08.21 & 0.38  & 4100-6700 & 1,2,4,5,8,9,12\\
1996.08.28 & 0.38  & 4100-6700 & 1,2,4,5,6,12\\
1996.08.29 & 0.38  & 4100-6700 & 3,7,8,9,10,11,13,14\\
1996.09.06 & 0.38  & 4100-6700 & 1,2,3,4,5,6,7,8,9,12,14\\
1996.09.07 & 0.38  & 4100-6700 & 5,10,13\\
1997.07.25 & 0.38  & 4100-6700 & 1,5\\
1997.07.26 & 0.38  & 4100-6700 & 2,3,4,6,7,8,9,10,11,12\\
1997.07.29 & 0.38  & 4100-6700 & 13,14\\
1997.07.30 & 0.38  & 4100-6700 & 11\\
1997.08.10 & 0.38  & 4100-6700 & 2,3,4,7,8,9,10\\
1997.08.11 & 0.38  & 4100-6700 & 1,5,6,11\\
1997.08.12 & 0.38  & 4100-6700 & 12\\
1997.08.13 & 0.38  & 4100-6700 & 1,4,7,11,12\\
1997.08.14 & 0.38  & 4100-6700 & 1,4,7,11,12\\
1997.08.15 & 0.38  & 4100-6700 & 11\\
1998.09.09 & 0.38  & 4500-9500 & 1,4,7,12\\
2000.09.12 & 0.38  & 4500-9500 & 2,3,6,8,13,14\\
2000.09.13 & 0.38  & 4500-9500 & 2,3,6,8,13,14\\
2000.09.14 & 0.38  & 4500-9500 & 2,3,6,8\\
2000.09.16 & 0.38  & 4500-9500 & 2,3,4,5,6,13,14\\
2000.09.17 & 0.38  & 4500-9500 & 1,2,6,7,14\\
2001.07.27 & 0.19  & 4500-9500 & 2,3,4,7,10\\
2001.07.27 & 0.38  & 4500-9500 & 6\\
2001.07.28 & 0.38  & 4500-9500 & 7,10,15\\
2001.07.28 & 0.19  & 4500-9500 & 2,3,4,6\\
2001.07.29 & 0.19  & 4500-9500 & 2,3,4,6\\
2001.07.29 & 0.38  & 4500-9500 & 7,15\\
2001.07.29 & 0.19  & 4500-9500 & 2,3,4,6\\
2001.07.30 & 0.19  & 4500-9500 & 1\\
2001.09.09 & 0.38  & 4500-9500 & 11,12\\
2001.09.09 & 0.19  & 4500-9500 & 1\\
2001.10.29 & 0.19  & 4500-9500 & 1,2,3,4\\
2001.11.02 & 0.19  & 4500-9500 & 1,4,6\\
2001.11.29 & 0.19  & 4500-9500 & 2,3\\
2001.11.30 & 0.19  & 4500-9500 & 1,3,7,10\\
2001.12.28 & 0.38  & 4500-9500 & 3,7,15,16\\
2002.01.02 & 0.38  & 4500-9500 & 7,15\\
2002.01.03 & 0.38  & 4500-9500 & 16\\
2002.01.04 & 0.38  & 4500-9500 & 3,7,16\\
2002.01.06 & 0.38  & 4500-9500 & 15,16\\
2003.05.21 & 0.19  & 4500-9500 & 1,5,16\\
2003.05.22 & 0.19  & 4500-9500 & 16\\
&&&\\[-4pt]
\hline
\end{tabular}
\label{j_obs}
\end{table}

Radial velocities from individual observations (hereafter referred to
as {\em epoch} radial velocities) of the program stars are given in Table~4.
For O and B type program stars they rest on individual measurement of HeI
and HeII lines. For the other, cooler program stars the radial velocities
come from measurement of the metallic absorptions lines (mainly Fe~I, Mg~I,
Ti~II). The radial velocities of the Be program star \#5 pertain to the
emission lines, which completely fill helium and hydrogen absorption lines.

\begin{figure}[!Ht]
\centerline{\psfig{file=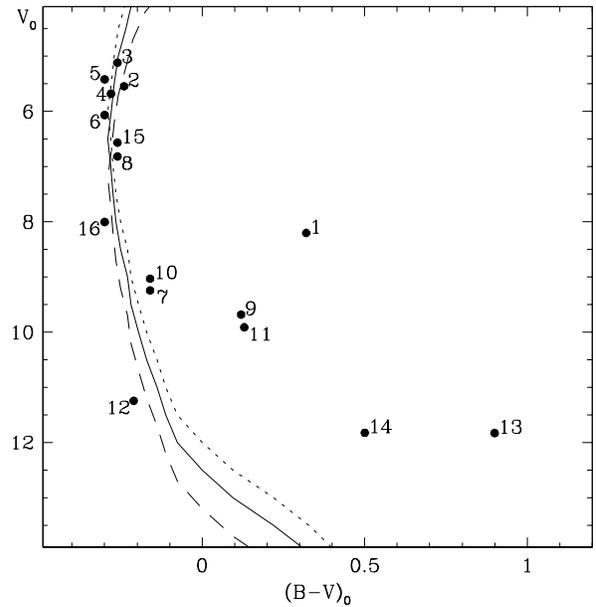,width=7.8cm}}
\caption[]{The program stars on the reddening corrected $V_\circ$,
$(B-V)_\circ$ diagram using $E_{B-V}$ from Table~3 and $V$, $B-V$ photometry
from Table~1 for $R_{\rm V}=A_{\rm V}/E_{\rm B-V}=$3.6 appropriate for
NGC~6913 according to Johnson (1962). The isochrone for solar metallicity
and 5 Myr is from Bertelli et al. (1994) and it is scaled to $m-M$=10.5
(dotted line), $m-M$=11.0 (solid line) and $m-M$=11.7 (dashed line). It is
evident how the cluster distance cannot be well constrained. In this paper
we adopt a 1.6~kpc distance.}
\label{zams}
\end{figure} 

	\subsection{Binaries and orbital solutions}

About half of the program stars have turned out to be spectroscopic
binaries. Table~5 summarizes the barycentric velocity, the membership and
the binary status based on epoch radial velocities in Table~4. Table~6 gives
the spectroscopic orbits computed for all the binary stars but \#6 and 16,
which are clearly binaries but the available radial velocities are not
enough to determine the orbital period and thus to allow to derive an
orbital solution. Therefore the RV$_{\odot}$ quoted for stars \#6 and 16 in
Table~5 is the mean of the measurements, not the barycentric velocity, and
the two velocities tend to differ with increasing eccentricity and paucity
of measurements. Consequently, the RV$_{\odot}$ of stars \#6 and 16 quoted
in Table~5 which differ by slightly more than 3$\sigma$ from the cluster
mean velocity cannot be considered as a firm indication that stars \#6 and
16 are field stars.

Similarly to Paper~I, the spectroscopic orbits have been obtained with a
Fortran code written by Roger F. Griffin (Cambridge University) and adapted
to run under GNU/Linux by us.

Program stars \#1--7 have been observed also by Liu et al. (\cite{liu89},
\cite{liu91}) who reported some epoch radial velocities for them. Such data
appear affected by large errors for the O and B stars (program stars
\#2--7), which make them useless in our analysis. They are instead in good
agreement with our velocities for star \#1, much cooler having a spectral
type F0~III. The reason for the poor quality of the Liu et al. radial
velocities of hot stars probably lies in the shortness of the wavelength 
range they
observed ($\sim$150~\AA) and by the fact that it is dominated by H$\delta$,
which we ignored in our analysis given the Balmer progression and its excessive
scatter compared to the much more performing HeI and HeII lines. It is also
worth noticing that Liu et al. did not recognized star \#7 as double lined,
in spite having observed it at orbital phase 0.66 when the velocity
separation between the components is $\sim$140~km~sec$^{-1}$ (cf. Figure~3)
and therefore outstanding.

The spectroscopic orbits of stars \#2 and 11 in Figure~3 and Table~6 are to
be considered quite preliminary, given their small amplitude, high
eccentricity and limited number of observations. Further observations are
obviously encouraged for these two stars. Photometric observations of the
double lined star \#7 are in progress to the aim of constraining the orbital
inclination and derive individual masses, and they will reported elsewhere
when completed.

\begin{table}
\begin{center}
\caption[]{Spectral types of the program stars from Wang \& Hu (\cite{wang}),
from photometry in the Vilnius system by Kazlauskas and Jasevicius (1986)
transformed by us into spectral types following Stray\v{z}is (1977), and
from classification of our Echelle spectra against the Yamashita et al.
(1977) reference spectral atlas. The last columns give the $E_{B-V}$ (from
Fitzgerald 1970 intrinsic colors) and the spectro-photometric distances for
our spectral classification and the photometry in Table~1.}
\begin{tabular}{rlcclc}
\hline
&&&&&\\[-5pt]
    &\multicolumn{3}{c}{spectral type}&\\ \cline{2-4}
\#  & WH00&Vilnius&our&$E_{B-V}$ & d (kpc) \\ 
&&&&&\\[-5pt]
      1  & F0 III  & F1 V  &  F0 III     & 0.11 & 0.2 \\ 
      2  & B0 I    & B1 Ia &  B0 I       & 0.97 & 2.0 \\
      3  & B0 I    & B1 Ia &  B0 I/II    & 1.06 & 1.4 \\
      4  & O7 II   & B1 Ia &  O9 Ib      & 0.92 & 2.3 \\
      5  & B0 IIIe &       &  B0 IIIe    & 1.07 & 1.2 \\
      6  & B0 II   & B2 Ia &  B0 V       & 0.93 & 1.1 \\
      7  & B4 II   & B3 V  &  B5 IV      & 0.26 & 1.6 \\
      8  & B0 III  & B0 III&  B0 Ib/II   & 0.99 & 3.0 \\
      9  & A4 V    & F0 V  &  A4 V       & 0.23 & 0.3 \\
     10  & B6 II   & B2 V  &  B5 V/IV    & 0.41 & 1.2 \\
     11  &         & A5 V  &  A5 III/II  & 0.39 & 1.9 \\
     12  &         & B2 V  &  B2 V       & 0.26 & 5.3 \\
     13  & G6 III  &       &  G5 III     & 0.00 & 1.1 \\
     14  & A8 II   & A4 V  &  F7 IV      & 0.00 & 0.7 \\
     15  & B1 II   & B3 Ib &  B1 III     & 1.38 & 1.6 \\
     16  &         & B1 II &  B0 V       & 0.83 & 2.8 \\
&&&\\[-5pt]
\hline
\end{tabular}
\end{center}
\label{classif}
\end{table}

\begin{table}[!t] 
\caption[]{Example of the Table containing the epoch radial velocities (and
their errors) for the program stars, available in full in electronic form
only.}
\centering
\begin{tabular}{crc}
\hline
&&\\
\multicolumn{3}{c}{\# 5}\\
&&\\
HJD & RV$_\odot$ & err \\
&&\\
50263.424 &$-$19.8 &1.5\\
50296.514 &$-$19.5 &1.2\\
50299.487 &$-$18.1 &0.9\\
50316.566 &$-$18.8 &0.9\\
50324.479 &$-$19.3 &0.9\\
50333.434 &$-$19.8 &0.9\\
50333.559 &$-$19.8 &0.9\\
50655.500 &$-$20.5 &0.9\\
50671.566 &$-$19.9 &1.7\\
51895.302 &$-$26.3 &0.9\\
52781.402 &$-$18.9 &1.0\\
&&\\
\hline
\end{tabular}
\label{tab_vel} 
\end{table}

        \subsection{Rotation velocities}\label{vrot}

Rotational velocities for the program stars are given in Table~5. They have
been derived from HeI lines for stars \#2, 3, 4, 6, 7, 8, 10, 12, 15, 16 and
FeI lines for the remaining ones, following the numerical relations for the
Asiago Echelle spectrograph calibrated in Paper~I (its Figure~7). No
rotation velocity is derived for the Be program star \#5 because all HeI
lines are badly affected by emissions. The correspondence of the rotational
velocity scale between HeI and FeI lines (which we have been forced to used
in all program stars with a spectral type later than B) has been carefully
checked on a grid of Kurucz rotationally broadened spectra we have
calculated on purpose.

From the spectral classification in Table~3, the stellar radii over the HR
diagram as tabulated by Strai\v{z}ys and Kuriliene (1981) and the observed
$V_{rot} \sin i$ projected rotation velocity, we have derived in the last
column of Table~5 the projected rotation period ($P_{rot}$/$\sin i$) for the
binaries with an orbital solution in Table~6. The projected rotation period
is obviously an upper limit to the true rotation period. Compared to the
orbital period in Table~6, it can be used to infer about the co-rotation
status of the binaries.

Star \#1 is an SB1 eclipsing binary and therefore the $\sin i$ projection
factor converge toward unity, which allows a direct comparison between
rotation and orbital periods, the former being twice longer. The lack of
synchronicity could be related to the primary evolving away from the main
sequence and the time scales of the two processes.

\begin{figure}[!Ht]
\resizebox{8cm}{!}{\includegraphics{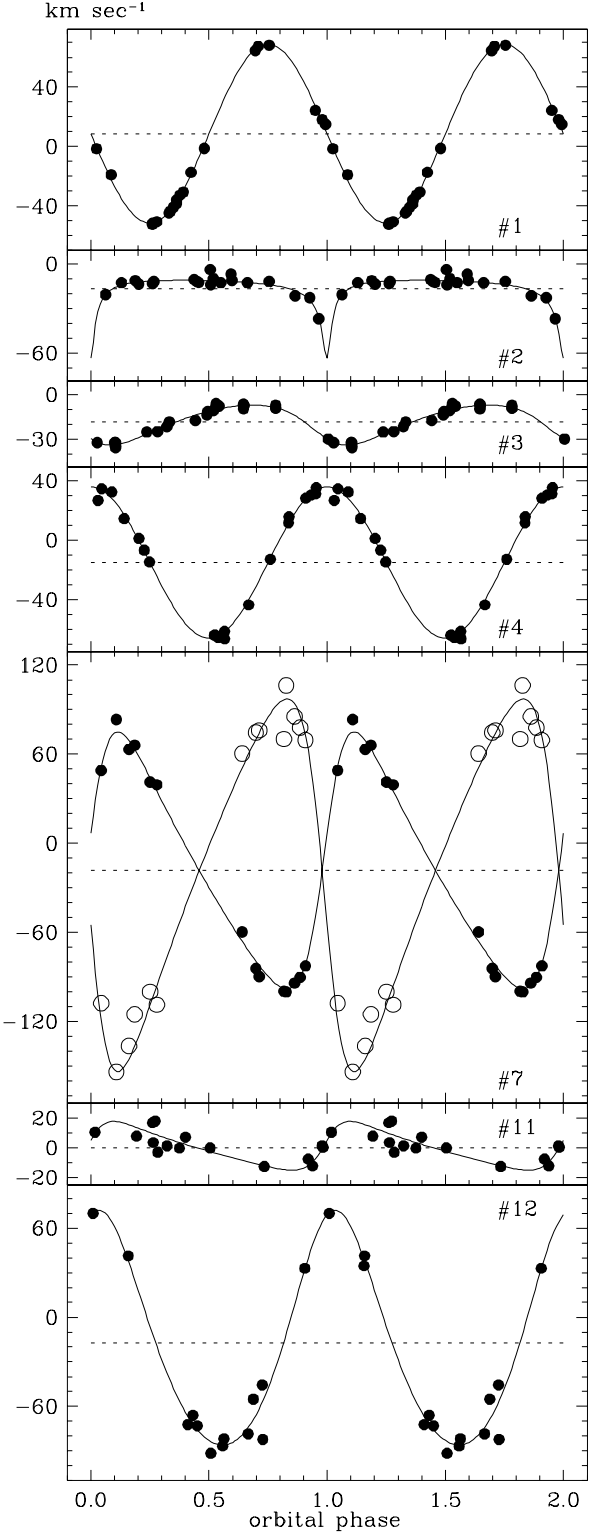}}
\caption[]{Orbital solutions for the binary program stars (cf. Table~6).}
\label{orbite} 
\end{figure}

\begin{table}[!Ht]
\caption[]{Heliocentric radial velocity (with its standard error) of the
program stars, binary status and cluster membership according to radial
velocities, and projected rotational velocity (with its standard error). The
radial velocity of the binary stars is the barycentric velocity from the
orbital solutions in Table~6. The last column gives the projected rotation
period $P_{rot}$/$\sin i$ of the solved binaries to be compared with the
orbital period.}
\begin{center}
\setlength{\tabcolsep}{3pt}
\begin{tabular}{rccccc}
\hline
&&&&\\
\#     & RV$_{\odot}$      & binary & member&V$_{rot}\sin i$& P$_{rot}$/$\sin i$\\
       & (km s$^{-1}$)     &        &       & (km s$^{-1}$) & (days) \\
\hline
&&&&\\
1      &   +8.6$\pm$0.5 & yes & no  & ~~26$\pm$1 & 5.0 \\	
2      &   --17$\pm$6   & yes & yes & 128$\pm$3  & 10 \\
3      & --18.5$\pm$0.6 & yes & yes & ~~84$\pm$2 & 15 \\
4      & --15.0$\pm$0.6 & yes & yes & 103$\pm$2  & 10 \\
5      & --19.4$\pm$0.3 & no  & yes &            &    \\
6      & --25.6$\pm$0.9 & yes &(yes)& 253$\pm$5  &    \\
7      &   --18$\pm$3   & yes & yes & ~~46$\pm$5 & 3.8\\
8      & --15.5$\pm$0.4 & no  & yes & ~~67$\pm$2 &    \\
9      & --5.6 $\pm$0.5 & no  & no  & ~~22$\pm$2 &    \\
10     & --14.8$\pm$0.7 & no  & yes & ~~64$\pm$7 &    \\
11     &      0$\pm$3   & yes & no  & ~~41$\pm$6 & 11 \\
12     &   --17$\pm$6   & yes & yes & 165$\pm$7  & 1.5\\
13     & --14.6$\pm$0.2 & no  & yes & ~~~0$\pm$2 &    \\
14     & --19.9$\pm$0.4 & no  & yes & ~~11$\pm$3 &    \\
15     &   --16$\pm$2   & no  & yes & ~115$\pm$4 &    \\
16     &   --27$\pm$4   & yes &(yes)&~~325$\pm$20&    \\
&&&&\\
\hline
\end{tabular}
\label{tab_bvel}
\end{center}
\end{table}

Given the masses estimated from the spectral type and the amplitude of
radial velocity variation, star \#7 probably has a high inclination too,
possibly being eclipsing itself. The rotational velocity in Table~5 pertains
to the B5~IV primary, the measurement of the secondary being too uncertain
given the difference in brightness. The rotation and orbital periods are
quite close, and in view of the uncertainties at play, the primary in star
\#7 looks synchronized.

Stars \#3 and 12 are evidently not co-rotating, because the projected
rotation period is at least several times shorter than the orbital period in
Table~6, and working on $\sin i$ can only enlarge the difference. For the
remaining binary stars \#2, 4 and 11, no conclusion can be drawn about the
co-rotation status, the projected rotation period being longer than the
orbital one.

	\subsection{Cluster membership}\label{rv_membership}

Sanders (\cite{sanders}, hereafter S73) has published an astrometric
investigation of 228 stars in the field of NGC~6913, identifying 92 possible
members. He has however used only one plate pair, with an epoch separation
of just 22~yr, with moreover the first epoch plate ``{\em severely blackened
by the moon}''. Consequently, noting the too large fraction of detected
members among the measured stars, he warned that the cluster separation from
the field is not satisfactory, and that the member/non-member status he
assigned may be frequently in error. S73 limiting magnitude is $V$=13.8, with
a completness limit not fainter than $V$=13.0 that corresponds to
1.2~M$_\odot$ on the main sequence of NGC~6913. Dias et al. (\cite{dias},
hereafter D02) have used proper motions from the Tycho-2 catalog to
accomplish the astrometric member segregation, following the analytical
approach of Sanders (1971). They have in common 24 stars with S73. As
Figure~4 shows, for 6 of the 24 common stars the membership status of S73
and D02 are in disagreement, and for the remaining 18 there is a fair
agreement.

\begin{table*}[!tH] 
\caption[]{Orbital solution for the binary stars we discovered in NGC 6913.
For binaries \#6 and 16 the epoch radial velocities in Table~4 did not
allowed a determination of the orbital period and therefore the derivation
of the orbit. The errors are given in parenthesis in units of the last
digit. The last raw gives the r.m.s. deviation of the solution from the observed
radial velocities.}
\begin{tabular}{rllllllll}
\hline
&&&&&&&\\
\multicolumn{2}{c}{orb. element} &\multicolumn{7}{c}{program star \#}\\ \cline{3-9}
                  && \multicolumn{1}{c}{1}
                   & \multicolumn{1}{c}{2}
                   & \multicolumn{1}{c}{3}
                   & \multicolumn{1}{c}{4}
                   & \multicolumn{1}{c}{7}
                   & \multicolumn{1}{c}{11}
                   & \multicolumn{1}{c}{12}\\
&&&&&&&\\
P \hfill    &(days)          & 2.70466(1) & 1.7075(1)   & 697(7)    & 3.51042(5) & 3.4588(1)     & 1.56842(4) & 4.0350(5)      \\
e           &                & 0.0        & 0.8(7)      & 0.21(5)   & 0.0        & 0.35(2)       & 0.4(2)     & 0.13(9)        \\ 
K           &(km sec$^{-1}$) & 60.1(7)    & 26(6)       & 13.5(6)   & 50.9(8)    & 87(2), 125(5) & 16(5)      & 79(6)          \\
$\gamma$    &(km sec$^{-1}$) &+8.6(5)     & --17(6)     & --18.5(6) & --15.0(6)  & --18(3)       &0(3)        & --17(6)         \\
T$_{0}$     &(+2450000)      &1132.626(4) & 1362.2(1)   & 652(25)   & 892.40(1)  & 923.96(6)     & 782.3(1)   & 668.7(3)        \\
$\omega$    & (deg)          & 0.0        & 187(14)     & 134(12)   & 0.0        & 282(7)        & 283(49)    & 344(26)          \\
a\,$\sin i$ &(10$^6$ km)     & 2.24(2)    & 0.4(17)     & 126(6)    & 2.45(4)    & 3.84(9)       & 0.3(1)     & 4.3(3)           \\
f(m)        &                & 0.061(2)   & 0.001(9)    & 0.16(2)   & 0.048(2)   & 0.19(1)       &0.0005(5)   & 0.20(4)         \\
r.m.s.      &(km sec$^{-1}$) &1.41        & 1.09        & 1.36      & 1.72       & 1.57          & 1.49       & 1.57          \\
\hline
\end{tabular}
\label{tab_orb} 
\end{table*}

A firm membership segregation is required for any kinematical investigation
of the cluster, and to achieve the best possible result, the astrometric
data should be complemented by radial velocities and placing of the program
stars on the HR diagram.

The radial velocity distribution of the program stars is presented in
Figure~5, where the cluster grouping at $-$16.9~km~sec$^{-1}$ is evident,
with a dispersion of $\sigma$=2.0~km~sec$^{-1}$.

Table~7 summarizes the membership status according to the astrometric
investigations of S73 and D02, the photometry presented and discussed by
Crawford et al. (1977) and Joshi et al. (1983), the spectral classification
of Wang and Hu (2000), our reddening free HR diagram of Figure~2 and the
radial velocities in Table~5. Their combination provide our final, adopted
membership reported in the last column of Table~7.

\begin{figure}[!Ht]
\resizebox{\hsize}{!}{\includegraphics{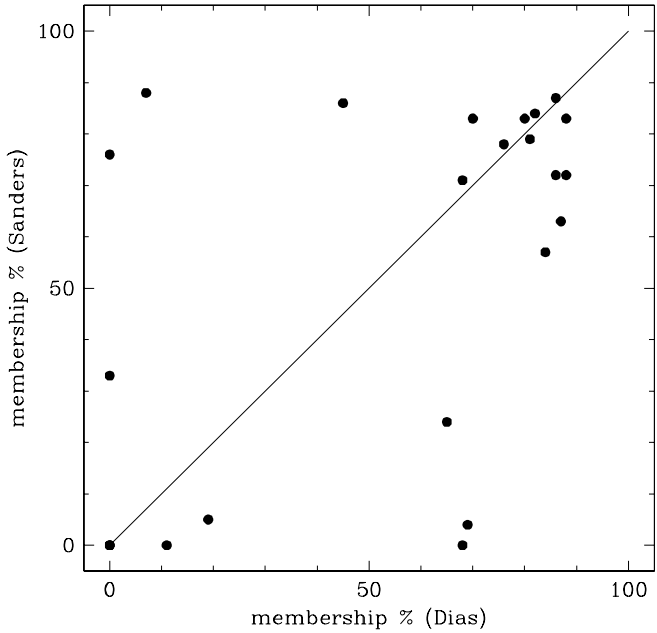}}
      \caption[]{Comparison between Dias et al. (\cite{dias}) and Sanders
(\cite{sanders}) membership data for the 24 stars in common.}
\label{dias_sanders_ppm}
\end{figure}
\begin{figure} [!Ht]
\resizebox{\hsize}{!}{\includegraphics{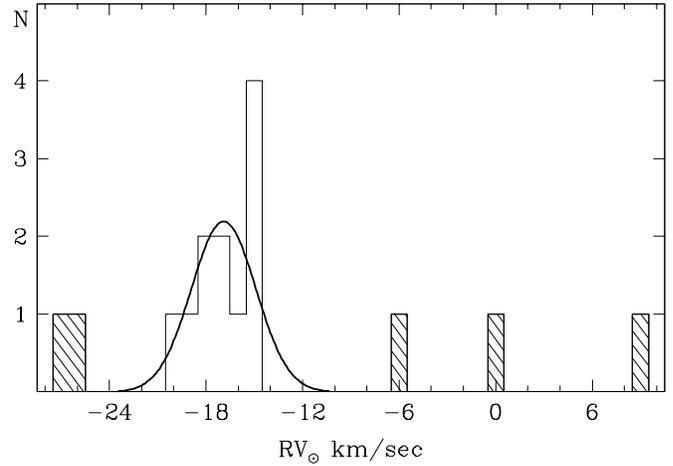}}
\caption[]{Radial velocity distribution of the program stars. A Gaussian with
center at RV$_\odot=-$16.9 km~sec$^{-1}$ and $\sigma$=2.0 km/sec fits the
member stars. Shaded elements represent stars which radial velocity differs
by more than 3$\sigma$ from the cluster mean velocity.}
\label{disp_RV} 
\end{figure}

	\section{Internal kinematic of NGC~6913}

       \subsection{A foreground cloud hiding part of the cluster}

As reviewed in the Introduction, all previous investigations of NGC~6913
agree on the large differential and total reddening affecting the cluster.
Inspecting the Palomar charts it is evident how NGC~6913 lies close to a
very thick interstellar cloud that seem to hide part of it.

Our and literature estimates about the cluster distance converge toward a
1.6 kpc value. Studies of the interstellar extinction toward NGC~6913
(Crawford et al. 1977, Neckle and Klare 1980) agree on a steep increase of
the extinction at 1 kpc, about halfway the distance to the cluster, as if a
single, major cloud is responsible for the majority of the extinction all
the way to NGC~6913.

In Figure~6 we compare the same field centered on NGC~6913 as seen on
Palomar POSS-II blue charts and by IRAS satellite at 100~$\mu$m. The
north-east quadrant is clearly deprived of stars in the optical image, while
it is bright in the far infrared, a clear sign of thick dust absorbing in
the optical and emitting in the IR. Star counts from USNO-A2.0 (stars
detected in both blue and red POSS-I prints) and near-infrared 2MASS survey
(stars detected both in $J$ as well as $H$ and $K$ bands) strongly support
the argument of a strong foreground interstellar extinction crossing the
field of NGC~6913 and increasing steeply toward the north-east quadrant.

We expect this foreground optically thick cloud to hide from view part of
the cluster, even if the cluster center seems confidently identifiable with
the grouping of the massive O and B stars seen in the optical.

	\subsection{Cluster mass}\label{cl_mass}

There are no published estimates of the NGC 6913 total mass. Two static
approaches are considered in this subsection, a dynamical one is
investigated in sect. 3.4.

A lower limit to the cluster mass is obtained by adding the mass appropriate
to the spectral type of known members (luminous mass). Taking Wang and Hu
(2000) spectral types of all S73 probable members classified by them and
calibration into masses from Strai\v{z}ys and Kuriliene (1981), it results
\begin{equation}
M_{cl}\simeq 700~M_{\odot}
\end{equation}
S73 warned that a sizeable fraction of his members could be spurious. We
assume here that they compensate for those fainter than the S73 completness
magnitude, and therefore 700~M$_{\odot}$ is taken as a fair indication of
the total luminous mass of the cluster.

Assuming that the members of NGC~6913 distribute according to the Salpeter
(1955) law $N(m)=C\, m^{-2.35}$ offers another possibility to estimate the
cluster mass. It seems fair to assume that all O and B type cluster members
have been detected and recognized as such in the Wang and Hu (2000) spectral
survey of NGC~6913. They are 31 in total, spanning the range between 6 and
67 $M_{\odot}$. This allows to estimate the constant $C$
\begin{equation}               
31=C\int_{6}^{67}m^{-2.35}dm\ \ \ \longrightarrow\ \ \ C=489
\end{equation}
which provides an initial total stellar mass for the cluster amounting to
\begin{equation}
M_{cl}=489\int_{0.08}^{120} N(m)\, m~dm\simeq 3100~M_{\odot}
\end{equation}
distributed in about $10^{4}$ member stars. This is an upper limit to
current cluster mass, because ($a$) stars became unbound early in the
cluster evolution when it started to lose gaseous mass blowed away by the
energetic winds of the first massive stars that formed, ($b$) the
relaxation mechanism leads to evaporation of the lighter members, and ($c$) 
the mass function appears to flatten toward lower masses (cf. Brice\~{n}o et al. 2002). The
Salpeter's and observed mass functions are compared in Figure~7. In sect.3.5
and Figure~8 will later show how there are possible evidences that the cluster
is at least partially relaxed. Supposing that the associated evaporation of
members affects stars fainter than S73 completness magnitude, it is found
that
\begin{equation}
M_{cl}=489\int_{1.2}^{67} N(m)\, m~dm\simeq 1000~M_{\odot}
\end{equation}
distributed in 350 members, comparable to the luminous mass in Eq.(1).

	\subsection{Tidal and half-mass radii}\label{sec_t_radius}

Cluster member venturing on orbits wider than the cluster tidal radius have
a fair chance to become unbound due to the action of the gravitational field
of the Galaxy. The cluster tidal radius is defined as (cf. Binney and
Tremaine 1987):
\begin{center} 
\begin{equation}\label{tidal_r}
R_{t}=R_G\Big(\frac{M_{cl}}{3M_G}\Big)^{\frac{1}{3}} 
\end{equation}
\end{center} 
\noindent
where $R_G$, $M_G$ are the radius of the cluster galactic orbit and the
galactic mass contained within ($R_G$=8.3 kpc and $M_G$=9.5 10$^{10}$
M$_\odot$). Inserting M$_{cl}$=700 M$_\odot$ provides R${_t} \sim$ 10 pc,
corresponding to 20$^\prime$ at the estimated 1.6 kpc cluster distance. On
photographs, the cluster tidal radius is conventionally taken as the
distance from the cluster center at which the stellar density drop to field
value, estimated by Lyng\aa\, (1987) to be $\sim 10^\prime$ for NGC~6913,
which corresponds to 4.7 pc at the cluster adopted 1.6~kpc distance. For
R${_t}$=4.7~pc Eq.(5) would provide a much lighter cluster mass,
M$_{cl}$=50~M$_\odot$. Eq.(5) describes a statistical relation of
equilibrium between cluster and galactic potentials that can be realized
only on times scales long enough for perturbation fluctuations to average.
Given the very young age of NGC~6913 (corresponding to just 2\% of its
galactic orbital period), the conditions supporting Eq.(5) can hardly be
matched, and a M$_{cl}$ estimated via Eq.(5) cannot be trusted. For the same
reasons, the $R_t$ estimated from star counts has to be preferred to the
value estimated using Eq.(5). 

\begin{table}
\caption[]{Cluster membership of program stars from literature according to
astrometric, photometric, and spectral type criteria, and ours based on
spectrophotometric parallaxes and radial velocities.  The last column
gives our final membership status obtained by merging the results of the
various criteria. {\sl S73} = Sanders (\cite{sanders}), {\sl D02} = Dias at
al. (\cite{dias}), {\sl C77} = Crawford et al. (1977), {\sl J83} = Joshi et
al. (1983), {\sl W00} = Wang and Hu (2000).}
\begin{center}
\setlength{\tabcolsep}{3pt}
\begin{tabular}{rccccccccc}
\hline
&&&&&\\[-5pt]
    &astrom.  &&photom.  && spctr&& ours\\ \cline{2-2}\cline{4-4}\cline{6-6}\cline{8-8}
\#  & S73 D02 && C77 J83 && W00    && dist. rv~~&& \\
&&&&&\\[-5pt]
1   & y ~~n && n ~~n && n&& n ~~n  && N \\ 
2   & y ~~y && y ~~y && y&& y ~~y  && Y \\
3   & y ~~y && y ~~y && y&& y ~~y  && Y \\
4   & y ~~y && y ~~y && y&& y ~~y  && Y \\
5   & y ~~y && y ~~y && y&& y ~~y  && Y \\
6   & y ~~y && y ~~y && y&& y ~~(y)&& Y \\
7   & y ~~y && n ~~y && y&& y ~~y  && Y \\
8   & y ~~y && y ~~y && y&& y ~~y  && Y \\
9   & n ~~n && y~~~~ &&  && n ~~n  && N \\
10  & y ~~y && n ~~y && y&& y ~~y  && Y \\
11  & n ~~n && n~~~~ &&  && y ~~n  && N \\
12  & n ~~n && y~~~~ &&  && n ~~y  &&(N)\\
13  & y ~~y && n ~~y && y&& y ~~y  && Y \\
14  & y ~~y && n ~~y && y&& n ~~y  && Y \\
15  & y ~~y && ~~~~n && y&& y ~~y  && Y \\
16  & ~~~~y &&       &&  && y ~~(y)&& Y \\
&&&&&&\\[-5pt]
\hline
\end{tabular}
\end{center}
\label{membership}
\end{table}

The half-mass radius (R$_{hm}$) is a useful quantity frequently used in
N-body simulations. We have estimated it by measuring the radius that
contain half of the 700 M$_\odot$ luminous mass of Eq.(1), which turned out 
to be 8$^\prime$, corresponding to R$_{hm}$=3.7 pc at distance of 1.6 kpc.
Following Binney and Tremaine (1987) the virial radius can be expressed as
\begin{center}
\begin{equation}
R_{vir}=\frac{R_{hm}}{0.4}\simeq 9{\rm ~pc}
\end{equation}
\end{center}

\begin{figure*}
\resizebox{\hsize}{!}{\includegraphics{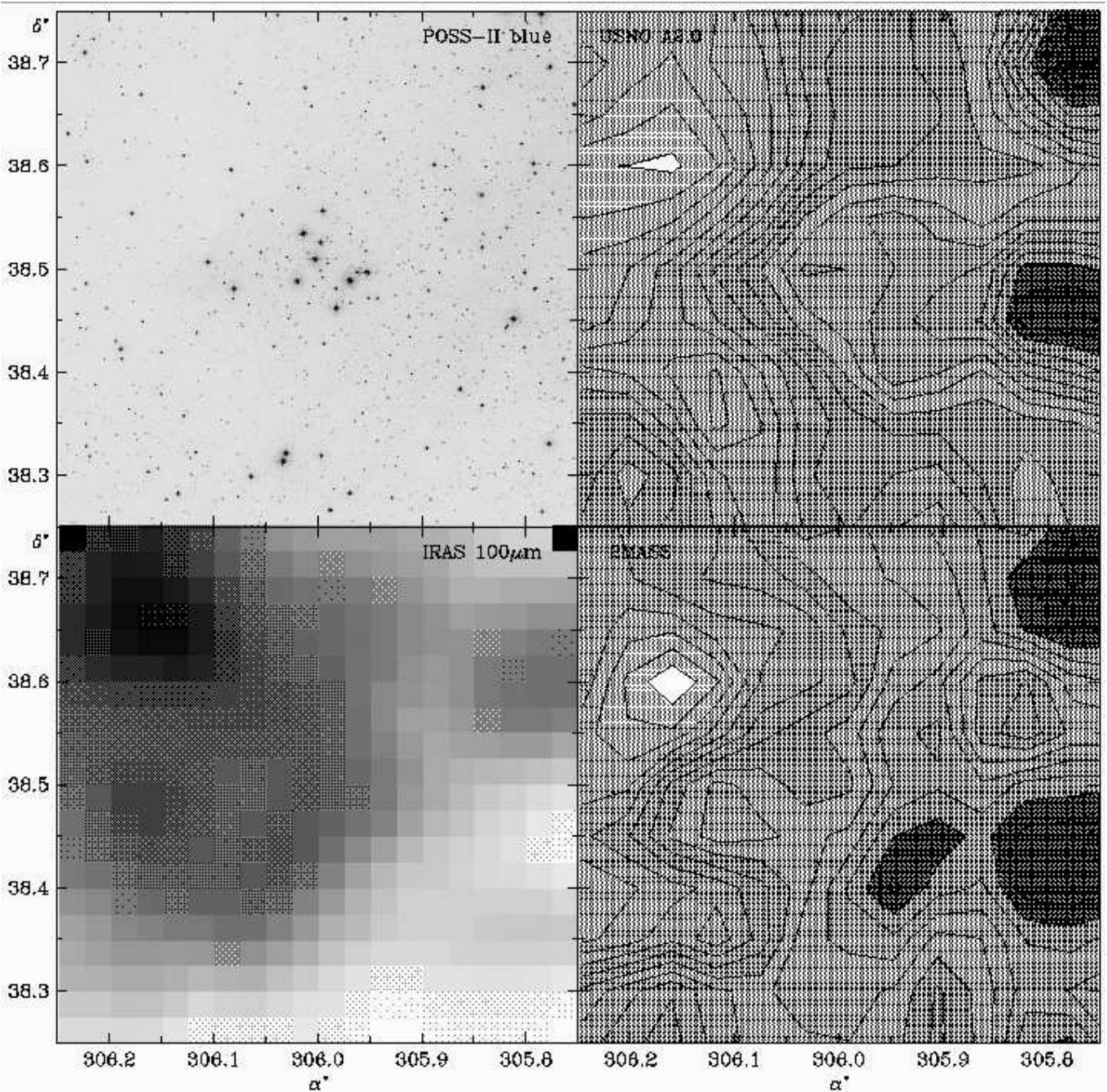}}
\caption[]{Comparison of a 30$^{\prime}\times$30$^{\prime}$ field from
POSS-II (top left) and IRAS 100 $\mu$m wavelength image (bottom left)
centered on NGC~6913. Stellar surface density of the same stellar field is
obtained from USNO-A2.0 optical catalog (top right) with minimum and maximum
density of 0.5 and 4.7 stars/arcmin$^2$ (clearest and darkest zone)
respectively and grey levels in step of 0.25 stars/arcmin$^2$, and from
2MASS infrared catalog (bottom right) with minimum and maximum density of
6.9 and 14.2 stars/arcmin$^2$ (clearest and darkest zone) respectively and
grey levels in step of 0.5 stars/arcmin$^2$.}
\label{ngc6913_iras}
\end{figure*}

	\subsection{Internal velocity dispersion and dynamical mass}

The mean radial velocity of NGC~6913 members is
\begin{equation}
RV^{cl}_{\odot}=-16.9\pm 0.6~{\rm km/sec}
\end{equation}
with an observed dispersion of $\sigma_{obs}$=2.0~km~sec$^{-1}$. The average
uncertainty of the radial velocity of individual members is
$\sigma_{instr}$=1.0~km~sec$^{-1}$, which leads to an intrinsic radial
velocity dispersion of NGC~6913 members amounting to
\begin{equation}
\sigma_{rv}=1.70~{\rm km~sec}^{-1}
\end{equation}
the three quantities being related as
$\sigma_{rv}=\sqrt{\sigma_{obs}^2-\sigma_{instr}^2}$.
Many of the physical quantities for NGC~6913 derived in the following
critically depend upon this number. We are confident that our observations,
reductions and measurements are state-of-the-art and involve so many high
resolution observations of such a large number of stars over a so long
period of time, that doing much better with 2m-class telescopes is hardly
feasible. On the other hand, the high competition to access larger
instruments prevents them to be assigned on programs like the present one that
require such a massive telescope usage (even if equipped with MOS devices).
Undoubtly, observations of more member stars over longer period of
times would be useful to better constrain the intrinsic radial velocity
dispersion, but this does not appear as an easy task. Therefore, we believe
that in absence of deeper/better efforts, our measurement of the intrinsic
radial velocity dispersion of NGC~6913 is worth some dynamical considerations 
that we develop in the rest of this paper.

The virial theorem links the cluster mass $M$ within the radius $R$ to the
velocity dispersion in the form
\begin{equation} 
\overline{v^2}=\frac{GM}{2R_{vir}}=3\sigma^2_{rv}
\end{equation}
For a virial radius of 9~pc, the virial cluster mass becomes
\begin{equation}
M_{tot}=\frac{6R_{vir}\sigma^{2}_{rv}}{G}\simeq 4 \, 10^4~M_{\odot} 
\end{equation}

There is a large difference between luminous and virial total mass for the
cluster (a factor of sixty, 7\,10$^2$ vs 4\,10$^4$~M$_\odot$). Could it be
intra-cluster mass ? Wang and Hu (2000) have derived a differential
reddening across the cluster amounting to $\Delta E_{B-V}$=1.82 mag. Even
supposing the effect is solely caused by intra-cluster extinction (from
above discussion the major contribution to differential reddening should
actually come from the optically thick, foreground dust cloud) the amount of
dust required to produce it is of the order of 4\,10$^3$~M$_\odot$, derived
assuming a simple spherical symmetry, constant density, a standard
dust-to-gas ratio ($N(H)=5.8\,10^{21}\,E_{B-V}$~atoms~cm$^{-2}$, Kilian
1992) and that a $\Delta E_{B-V}$=1.82 mag differential extinction is
accumulated as result of the line of sight crossing the whole cluster.
Even under these assumptions the corresponding amount of intracluster material
is far from filling the gap between the luminous (i.e. Eq.1) and dynamical
mass (i.e. Eq.10).

\begin{figure} [!t]
\resizebox{\hsize}{!}{\includegraphics{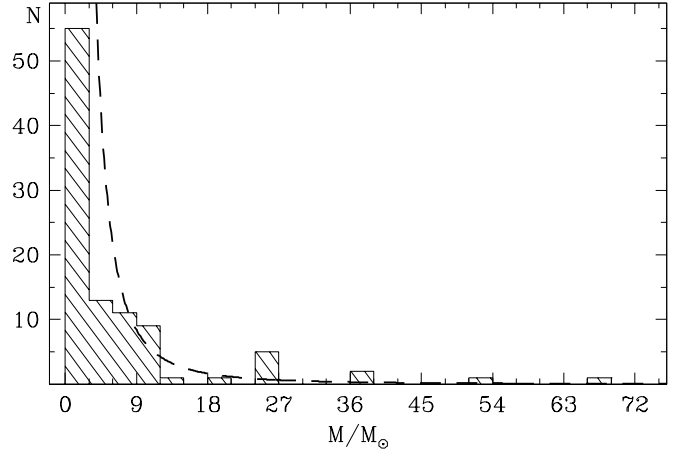}}
\caption[]{Mass function of the cluster NGC~6913 as it results from the
spectral classification of the member stars. The dashed line is a fit with a
Salpeter law $N(m)=489\, m^{-2.35}$ to O and B stars ($M\geq$6~M$_\odot$).}
\label{F_massa} 
\end{figure}

Possible explanations to solve such a discrepancy 
could be any combination of the following effects:\\
({\em i}) the large and optically thick interstellar cloud discussed in
sect. 3.1 hides from view a significant portion of the cluster, causing an
underestimate of the tidal radius and luminous mass from star counts.
Deep infrared imaging at $K$ band and longer wavelengths could test
this scenario;\\
({\em ii}) the cluster is still relaxing and some of the stars considered as
members are actually unbound leaving the cluster at velocities just larger
than the escape one, inflating the apparent dispersion of radial velocities.
An example could be star \#6 which is a member according to both astrometric
investigations and combined photometric+spectroscopic criteria, and lies
projected close to cluster center sporting one of the earliest spectral
types (B0~V). Its radial velocity is well determined in $-25.6$($\pm
0.9$)~km~sec$^{-1}$, which is however 8.7~km~sec$^{-1}$ away from the cluster mean velocity
of $-16.9$($\pm 0.6$)~km~sec$^{-1}$. The difference exceeds 4$\sigma_{obs}$.
The escape velocity from NGC~6913 is
\begin{equation}
v_{esc}=\sqrt{12\sigma^2_{rv}}=5.9\ \ {\rm km~sec^{-1}}
\end{equation}
supporting the possibility that star \#6 is actually escaping. Kroupa et al.
(2001) N-body simulations of young clusters confirm that the observed
dispersion of radial velocities could be biased by unbound members;\\ 
({\em iii}) some cluster members could be components of very wide binaries,
which orbital motion amount to just a few km~sec$^{-1}$ and cannot be detected
by observations spanning less than some decades;\\
({\em iv}) a cluster angular moment larger than zero, with the axis pointing away
from the line of sight, could pass unnoticed given the restricted number of
observed stars and still contribute a fraction of the observed radial
velocity dispersion. A cluster rotation having a projected radial component
of $\sim$1.5~km~sec$^{-1}$ at half mass radius could be enough to bring the
virial mass in agreement with the luminous mass.

\begin{figure}[!t]
\resizebox{\hsize}{!}{\includegraphics{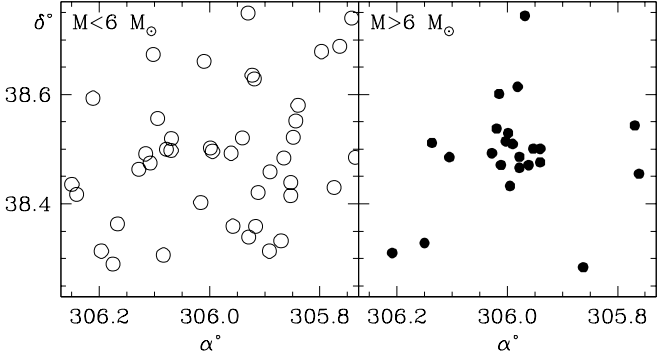}}
      \caption[]{Member stars of NGC~6913 with mass smaller (left) and
larger (right) then 6 M$_{\odot}$. Massive stars show a clear clustering
while the lighter seems randomly distributed.}
       \label{segr_masse}
   \end{figure}

\subsection{Evaporation of members}

Lighter cluster members seem evaporating from NGC~6913 as the comparison
in Figure~7 between the observed and Salpeter mass function supports,
and as mass segregation in Figure~8 suggests.

In fact, for O and B type members the mean mass is $\overline{m}\simeq 22$ M$_{\odot}$
and being $\sigma_{rv}=1.70$~km~sec$^{-1}$ the radial velocity intrinsic dispersion,
the mean kinetic energy of O and B type members is 
\begin{equation}
\overline{E_{c}}=\frac{1}{2}\overline{m}v^{2}=\frac{3}{2}\overline{m}\sigma^{2}_{rv}=1.9\,10^{45}\mbox{erg}
\end{equation}
If the cluster is relaxing toward energy equipartition, this is the mean kinetic energy 
to which its members are aiming, and there must be a value of the mass below which the velocity
exceeds the escape value and the stars tend to evaporate. Such limiting mass $m_{lim}$ follows from
\begin{equation}
\frac{1}{2}m_{lim}v_{esc}^{2}=\overline{E_{c}}=1.9\,10^{45}\mbox{erg}\ \ 
              \longrightarrow\ \ m_{lim}=5.5~{\rm M}_\odot
\end{equation}
However, this is actually an upper limit to the mass below which the members
tend to leave the cluster due to evaporation. In fact, if the cluster core
seems already relaxing in spite of the very young age, this could not be yet
the case for the lower density outer regions of the cluster, with the
relaxation being a process moving outward from the cluster center on
timescales longer than the NGC~6913 age. Only devoted, deep photometric
investigations can address what is the turning mass in NGC~6913 for which
lighter members are already experiencing evaporation and what is its radial
dependence.

	\subsection{Crossing time, relaxing time, mass segregation}

Crossing time and relaxing time are theoretical quantities which play a
relevant role in cluster dynamics. In fact they are closely related to the
mass of the cluster, its dynamical status and the number of members.
The crossing time is related to virial radius and cluster mass by ($v$ from Eq.7)
\begin{equation}
t_{cr}=\frac{2 R_{vir}}{v}=\sqrt{\frac{8R_{vir}^{3}}{GM_{cl}}}
\end{equation}
and the relaxing time relates to number of members, cluster mass and half mass radius as
\begin{center}
\begin{equation}\label{t_relax}
t_{relax}=\sqrt{\frac{R_{hm}^{3}}{GM_{cl}}}\,\frac{N}{8\log{(0.4N)}}
\end{equation}
\end{center}
(from Spitzer \cite{spitzer}, derived for the case of globular clusters with
similar mass members).

In the case of the luminous mass, $M_{cl} \sim$700~M$_\odot$ and $N$=92 (cf
setc. 2.4), it is $t_{cr}$=45 Myr and $t_{relax}$=30 Myr, significantly
longer that the estimated cluster age ($\sim$5~Myr). For the Salpeter's mass
$M_{cl} \sim$3100~M$_\odot$ and $N$=10$^4$ (cf. Eq. 3), the crossing time
reduces to $t_{cr}$=20 Myr while the relaxation time goes up to
$t_{relax}$=650~Myr, both still larger than cluster lifetime.

NGC~6913 appears relaxed, at least in its core (where the relaxation time is
expected to be shorter given the higher mass density) as mass segregation in
Figure~8 shows. It is worth to note that concentration of massive stars
toward cluster center is observed in some clusters to be present since their
birth and not as a the result of purely dynamical evolution (i.e. the Orion
trapezium system, see Hillenbrand \& Hartmann, \cite{hillenbrand}) and that
some N-body simulations (Portegies Zwart et al. \cite{portegies}) show mass
segregation to happen in clusters over ages shorter then the canonical
t$_{relax}$. So, the apparent concentration of heavier members toward
the center of NGC~6913 could also be due to mechanism other than dynamical
relaxation.

	\subsection{Binaries}

The dispersion of velocities in NGC 6913 is
$\sigma=\sqrt{3}\sigma_{rv}=$2.9~km~sec$^{-1}$. Binaries with orbital
velocities faster than it tend to survive close encounters, those orbiting
slower risk ionization (Kroupa 2000). An orbital velocity of
2.9~km~sec$^{-1}$ corresponds to an orbital period of $10^{6.3}$ days for a
binary with a total mass of 5.5~M$_\odot$, which would go clearly undetected
in the course of a 6-yr long monitoring like ours. Such a binary would have
an angular separation of 0.4~arcsec (within detection threshold of current
observational techniques from the ground) which would rise to 4.4 arcsec for
members with a total mass of 67~M$_\odot$ and decrease to 0.07~arcsec for
members with a total mass of 1~M$_\odot$.

All binaries detected in this investigation appear strongly bound, not
ionizable by close encounters with other cluster members, and quite probably
primitive (the short cluster age argue against a capture scenario). Their
large eccentricities and non synchronous orbits indicates how far they still
are from tidal circularization of the orbits and locking of the rotation and
orbital periods which characterize the field binaries. Wider, more ionizable
binaries are beyond the realm of spectroscopy, and would be profitably
searched for by high spatial resolution imaging.

	\section{Desiderata}

A deep and wide field photometric investigation of NGC~6913 and surrounding
field would be a good starting point to better constraint the total mass, tidal
radius and drop in the luminosity function of the cluster, and to address
the large discrepancy between observed luminous mass, integrated IMF mass
and the observed virial mass. At the cluster distance and reddening, UBVRI
photometry complete to $V=20$ will map all cluster members more massive than
0.8~M$_\odot$, thus venturing well into the realm of masses that should be already
evaporating from the cluster. Such a photometric investigation, which is
highly encouraged, should extend over a radius of not less that 20$^\prime$
from the cluster center and should include JHKL bands to overcome the
very strong differential extinction caused by the foreground interstellar cloud
discussed in sect. 3.1. Once members lighter than those here investigated
will be firmly identified, a study of their radial velocity
distribution would add a good deal of constraints to the
kinematical status and evolution scenario of NGC~6913.

\begin{acknowledgements}
We would like to thank R.Griffin for the provided software and P.M.Marrese for securing
the four spectra obtained during 2003.
CB has been finacially supported by ASI~I-R-050/02 and I-R-117/01 grants.
\end{acknowledgements}

\setcounter{table}{4}
\begin{table*}[t] 
\includegraphics{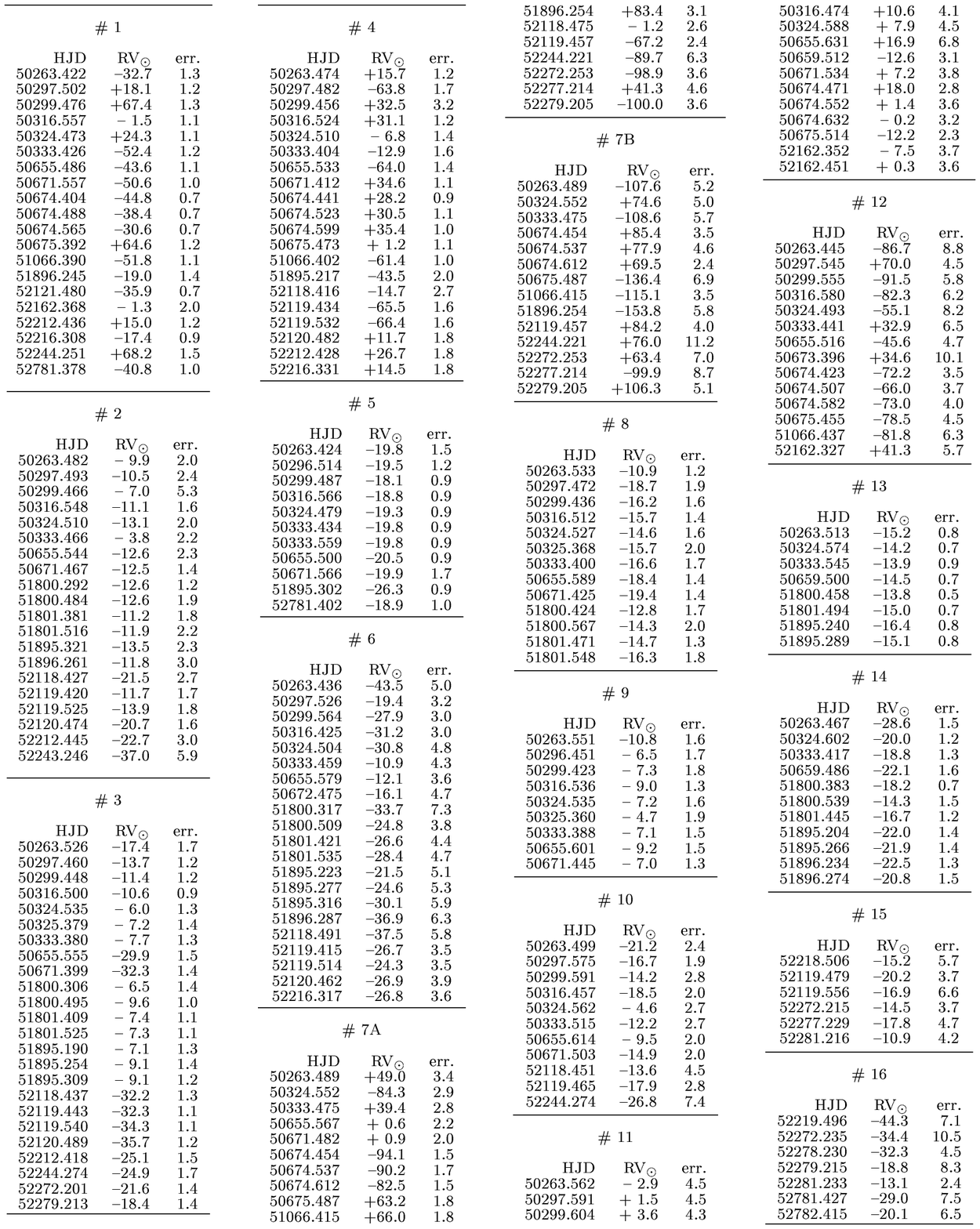}
      \caption[]{Epoch radial velocities of the program stars (full table given in electronic form only).}
\label{tab_vel} 
\end{table*}

\end{document}